\documentclass[preprintnumbers,superscriptaddress,showkeys,byrevtex]{revtex4}
\usepackage[T1]{fontenc}
\usepackage{amsmath,amsfonts,amssymb,amscd,amsxtra,amsthm}
\usepackage{graphicx}
\usepackage{epstopdf}
\usepackage{bm}
\usepackage{orcidlink}
\usepackage{stackengine}
\usepackage{subcaption}
\usepackage{placeins}
\usepackage{float}
\usepackage{pgfplots}
\usepgflibrary{plotmarks}
\begin{document}
%-------------------------------------------------
\preprint{PKNU-NuHaTh-2026}
%-------------------------------------------------
\title{A Maximum-Entropy Method for Zero-Skewness Valence GPDs\\ Constrained by Nucleon Electromagnetic Form Factors}
%-------------------------------------------------
\author{Seung-il Nam\orcidlink{0000-0001-9603-9775}}
\email{sinam@pknu.ac.kr}
\affiliation{Department of Physics, Pukyong National University (PKNU), Busan 48513, Korea}
%-------------------------------------------------
\date{\today}
%-------------------------------------------------
\begin{abstract}
We formulate a reduced-profile maximum-entropy method (MEM) framework for constructing constrained zero-skewness valence-quark generalized parton distribution (GPD) transverse profiles from the four nucleon electromagnetic form factors $F_1^p(t)$, $F_1^n(t)$, $F_2^p(t)$, and $F_2^n(t)$.  The form-factor sum rules fix only $x$-integrated moments of the GPDs; the forward limit of $H_v^q$ is fixed separately by the valence parton distribution functions, and the normalization of $E_v^q$ by the flavor anomalous magnetic moments.  These complementary constraints are combined through the ansatz $H_v^q(x,t)=q_v(x)\exp[t f_H^q(x)]$ and $E_v^q(x,t)=e_v^q(x)\exp[t f_E^q(x)]$, where the positive profile functions encode the $x$-dependent transverse structure.  Rather than attempting an unrestricted functional inversion, we use the entropy functional as a regularizing criterion on a low-dimensional positive profile manifold.  In the numerical proof-of-concept calculation, a smooth elastic form-factor input and analytic forward distributions are adopted, together with the reduced form $f(x)=0.05+(1-x)^2\exp(c_0+c_1x+c_2x^2)$, which suppresses local modes that elastic moments alone cannot constrain.  Within this reduced ansatz, the resulting profiles reproduce the imposed elastic moment constraints, satisfy the forward normalizations after discrete-grid normalization, and give impact-parameter distributions with the expected transverse shrinkage at large $x$.  The construction provides a controlled zero-skewness baseline for connecting elastic form-factor constraints to $x$-dependent transverse profiles, and it offers a stable starting point for future analyses incorporating empirical form-factor fits, modern PDF inputs, lattice-QCD generalized form factors, and hard exclusive observables.
\end{abstract}
%-------------------------------------------------
%-------------------------------------------------
\maketitle
%--------------------------------------------------
\section{Introduction}
%--------------------------------------------------
The spatial and momentum structure of the nucleon is encoded in several classes of observables, each of which projects the underlying QCD dynamics in a different way.  Elastic electromagnetic form factors (FFs) describe how the nucleon responds coherently to an external electromagnetic current.  They are functions of the invariant momentum transfer $t=(p'-p)^2$ and, in the spacelike region, are usually expressed in terms of $Q^2=-t>0$.  The Dirac and Pauli form factors $F_1^N(t)$ and $F_2^N(t)$ characterize the non-spin-flip and spin-flip components of the nucleon current, while the Sachs form factors $G_E^N(Q^2)$ and $G_M^N(Q^2)$ are directly connected to the electric and magnetic response in the Breit-frame interpretation.

Parton distribution functions (PDFs), in contrast, describe the distribution of longitudinal momentum among quarks and gluons in the infinite-momentum frame.  Ordinary PDFs are forward matrix elements and depend on the parton momentum fraction $x$ and a factorization scale.  They contain detailed information on the nucleon's longitudinal structure, but they do not directly resolve its transverse spatial distribution.  Thus, FFs and PDFs provide complementary but incomplete projections of the same underlying hadronic structure.

Generalized parton distributions (GPDs) unify these two descriptions.  They depend simultaneously on the average parton momentum fraction $x$, the skewness $\xi$, and the invariant momentum transfer $t$ \cite{Muller:1994ses,Ji:1996ek,Radyushkin:1996nd,Radyushkin:1997ki,Goeke:2001tz,Diehl:2003ny}.  In the forward limit, GPDs reduce to ordinary PDFs, while their Mellin moments yield form factors of local QCD operators~\cite{Guidal:2013rya,Kumericki:2016ehc,Kroll:2012sm}.  In particular, for unpolarized quark GPDs, $H^q(x,\xi,t)$ and $E^q(x,\xi,t)$, one has
\begin{equation}
H^q(x,0,0)=q(x),\qquad \int_{-1}^{1} dx\,H^q(x,\xi,t)=F_1^q(t),\qquad \int_{-1}^{1} dx\,E^q(x,\xi,t)=F_2^q(t).
\label{eq:intro_basic_limits}
\end{equation}
These relations show that GPDs contain both the PDF and FF information, and that form factors are not independent of the partonic description: they are $x$-integrated projections of GPDs.

The physical significance of GPDs is particularly transparent at zero skewness.  In this case, the Fourier transform with respect to the transverse momentum transfer yields impact-parameter-dependent parton distributions, which describe the transverse spatial distribution of partons with a fixed longitudinal momentum fraction $x$ \cite{Burkardt:2000za,Diehl:2003ny}.  The $t$-dependence of $H_v^q(x,t)$ therefore carries information on the $x$-dependent transverse size of the valence quark distribution.  Form factors, being $x$-integrated quantities, encode only the transverse distribution after summing over all parton momentum fractions.  Recovering the $x$-resolved transverse structure from form factors is therefore an \textit{inverse problem}.

This inverse problem is severely underdetermined if only a single form factor is used.  For example, knowing $F_1^p(t)$ gives only a charge-weighted integral of $H_v^u$ and $H_v^d$.  Infinitely many functions of $x$ can have the same integral.  However, the situation improves if several form factors are used simultaneously.  The proton and neutron Dirac form factors constrain different flavor combinations of $H_v^u$ and $H_v^d$, while the proton and neutron Pauli form factors constrain the corresponding spin-flip distributions $E_v^u$ and $E_v^d$.  Four form factors thus produce a coupled system of integral constraints.  When supplemented by forward PDF information, anomalous magnetic moment normalizations, and physically motivated profile priors, this system can be regularized.

The purpose of the present work is to formulate such a regularized construction scheme using the maximum entropy method (MEM).  MEM is a Bayesian regularization technique that selects the least-biased function compatible with known constraints and a default model \cite{Skilling:1984,Jarrell:1996rrw}.  MEM cannot create information that is absent from the form factors; what it offers is a controlled way to pick smooth transverse profiles $f_H^q(x)$ and $f_E^q(x)$ that reproduce the form-factor constraints without introducing structure beyond a physically motivated default profile.  The central ansatz of this work is $H_v^q(x,t)=q_v(x)\exp[t f_H^q(x)]$ and $E_v^q(x,t)=e_v^q(x)\exp[t f_E^q(x)]$, where $q_v(x)$ is the valence PDF and $e_v^q(x)$ is the forward limit of the Pauli-type GPD, normalized to the flavor anomalous magnetic moment.  The unknown functions $f_H^q(x)$ and $f_E^q(x)$ are transverse profile functions; since spacelike momentum transfer corresponds to $t<0$, positive profiles lead naturally to decreasing form factors with increasing $Q^2$.  The construction is thereby reduced to determining four profile functions within a restricted ansatz: $f_H^u(x)$, $f_H^d(x)$, $f_E^u(x)$, and $f_E^d(x)$.  Because the elastic form factors fix only the lowest Mellin moments of the GPDs and leave arbitrary local $x$-dependent structure unconstrained, this should be read as a baseline construction: within the reduced positive profile manifold, the entropy functional picks out a stable, least-biased solution compatible with the imposed moment constraints, not a global phenomenological extraction.  We emphasize that the novelty of the present work lies not in the exponential $t$-profile ansatz itself, which is standard \cite{Diehl:2003ny,Burkardt:2000za}, but in casting the four-form-factor problem as a single coupled MEM pipeline that combines the elastic moment constraints, the forward-PDF and anomalous-moment normalizations, and an entropy-based positivity-preserving regularizer.  This pipeline is deliberately input-agnostic: the elastic input, the forward PDFs, and the Pauli-forward shape enter only as replaceable modules, so the same reconstruction machinery applies without modification to different form-factor parametrizations and PDF sets.  The analytic inputs used below therefore serve to expose the structure of the inverse problem, and are not part of the method itself.

As a numerical demonstration, we implement the construction for $0\le Q^2\le 5~{\rm GeV}^2$ using a smooth baseline elastic form-factor input and analytic valence forward distributions.  The resulting GPD moments reproduce the input Sachs form factors at the percent level and satisfy the required forward normalizations associated with the valence quark numbers and the flavor anomalous magnetic moments after discrete-grid normalization. The transverse profiles are positive and decrease toward large $x$, leading to an $x$-dependent impact-parameter distribution in which large-$x$ partons are more localized near the transverse center of the nucleon.  We also examine the stability of the reduced-profile solution under variations of the entropy weight.  As an additional qualitative check, a simple beam-spin-asymmetry proxy is compared with the leading azimuthal pattern observed in CLAS12 data.  These numerical tests show that elastic form factors and forward parton information can define a physically motivated low-dimensional transverse profile within a maximum-entropy framework, while the result remains conditional on the chosen reduced profile manifold.

The rest of this paper is organized as follows. Sec.~\ref{sec:gpd_sum_rules} reviews the GPD sum rules and flavor decomposition of nucleon form factors.  Sec.~\ref{sec:ff_inputs} defines the elastic form-factor input used in the numerical implementation and briefly discusses alternative parametrizations.  Sec.~\ref{sec:pdf_input} introduces the forward PDF parametrization and the Pauli-forward input $e_v^q(x)$.  Sec.~\ref{sec:mem_formulation} formulates the coupled MEM reconstruction.  Sec.~\ref{sec:numerical_scheme} describes the discretized numerical implementation.  Sec.~\ref{sec:interpretation} discusses the physical interpretation of the reconstructed profiles and the consistency checks.  Sec.~\ref{sec:numerical_results} presents the numerical results, including the stability analysis, form-factor reproduction, impact-parameter distributions, and a qualitative beam-spin-asymmetry proxy.  Sec.~\ref{sec:summary} summarizes the method and outlines future extensions.

%--------------------------------------------------
\section{GPD sum rules and flavor decomposition}
\label{sec:gpd_sum_rules}
%--------------------------------------------------
We consider the zero-skewness limit as the first step toward a form-factor-based reconstruction.  This limit is appropriate for connecting GPDs to impact-parameter distributions and isolates the transverse momentum-transfer dependence without introducing the additional complexity of nonzero longitudinal momentum transfer.  The full GPDs $H^q(x,\xi,t)$ and $E^q(x,\xi,t)$ are accordingly approximated by $H^q(x,0,t)$ and $E^q(x,0,t)$.  The valence combinations for $0<x<1$ are defined by
\begin{equation}
H_v^q(x,t)=H^q(x,0,t)+H^q(-x,0,t),\qquad E_v^q(x,t)=E^q(x,0,t)+E^q(-x,0,t).
\label{eq:valence_def}
\end{equation}
The signs and conventions follow the standard definition in which negative $x$ represents antiquark contributions, $H^q(-x,0,t)=-H^{\bar q}(x,0,t)$, so that in the forward limit $H_v^q(x,0)=q(x)-\bar q(x)$ reduces to the ordinary valence PDF; the analogous relation holds for $E_v^q$.  In the valence sector, the above combinations are the natural objects whose first moments reproduce the flavor form factors.

The quark-flavor Dirac and Pauli form factors are obtained from
\begin{equation}
F_1^q(t)=\int_0^1 dx\,H_v^q(x,t),\qquad F_2^q(t)=\int_0^1 dx\,E_v^q(x,t).
\label{eq:F1F2_flavor_sum_rules}
\end{equation}
These equations are the basic moment constraints used throughout this work.  They are exact at the level of the GPD sum rules and do not depend on the skewness parameter.  The $\xi$-independence follows from polynomiality: the zeroth moment of the vector GPD corresponds to the local vector current, whose matrix element is parametrized only by the Dirac and Pauli form factors \cite{Ji:1996ek,Diehl:2003ny}.

The experimentally measured proton and neutron form factors are related to the flavor form factors through charge weighting.  Neglecting strange and heavier quark contributions and assuming isospin symmetry, one obtains
\begin{equation}
F_i^p(t)=\frac{2}{3}F_i^u(t)-\frac{1}{3}F_i^d(t),\qquad F_i^n(t)=\frac{2}{3}F_i^d(t)-\frac{1}{3}F_i^u(t),\qquad i=1,2.
\label{eq:proton_neutron_flavor}
\end{equation}
The inverse transformation is
\begin{equation}
F_i^u(t)=2F_i^p(t)+F_i^n(t),\qquad F_i^d(t)=F_i^p(t)+2F_i^n(t),\qquad i=1,2.
\label{eq:flavor_inverse_relation}
\end{equation}
Combining proton and neutron form factors in this way separates the $u$- and $d$-flavor contributions, which is a central advantage of treating the four form factors as a coupled system rather than fitting them separately.

At $t=0$, the normalization of the Dirac form factors is fixed by the valence quark numbers,
\begin{equation}
F_1^u(0)=2,\qquad F_1^d(0)=1.
\label{eq:F1_norm}
\end{equation}
The Pauli form factors at the origin are fixed by the flavor anomalous magnetic moments,
\begin{equation}
F_2^u(0)=\kappa_u,\qquad F_2^d(0)=\kappa_d,\qquad \kappa_u=2\kappa_p+\kappa_n,\qquad \kappa_d=\kappa_p+2\kappa_n.
\label{eq:kappa_flavor}
\end{equation}
Using $\mu_p=2.793$, $\mu_n=-1.913$, $\kappa_p=\mu_p-1$, and $\kappa_n=\mu_n$, we find
\begin{equation}
\kappa_u=1.673,\qquad \kappa_d=-2.033.
\label{eq:kappa_values}
\end{equation}
These numbers are imposed as normalization constraints on the forward Pauli-type functions $e_v^q(x)$.

%--------------------------------------------------
\section{Electromagnetic form-factor input}
\label{sec:ff_inputs}
%--------------------------------------------------
The reduced-profile MEM fit requires numerical input for the four nucleon electromagnetic form factors $F_1^p$, $F_1^n$, $F_2^p$, and $F_2^n$.  Since empirical analyses often provide the Sachs form factors, we first convert $G_E^N$ and $G_M^N$ into the Dirac and Pauli form factors.  With $Q^2=-t>0$ and $\tau=Q^2/(4M_N^2)$, the Sachs and Dirac-Pauli form factors satisfy
\begin{equation}
G_E^N(Q^2)=F_1^N(Q^2)-\tau F_2^N(Q^2),\qquad G_M^N(Q^2)=F_1^N(Q^2)+F_2^N(Q^2),
\label{eq:sachs_def}
\end{equation}
and the inverse relations are
\begin{equation}
F_1^N(Q^2)=\frac{G_E^N(Q^2)+\tau G_M^N(Q^2)}{1+\tau},\qquad F_2^N(Q^2)=\frac{G_M^N(Q^2)-G_E^N(Q^2)}{1+\tau}.
\label{eq:sachs_to_dirac_pauli}
\end{equation}

In the present proof-of-concept calculation, we use a smooth baseline parametrization of the elastic form factors.  Its role is methodological: it provides a controlled input to test whether the MEM framework can construct zero-skewness valence GPD profiles that reproduce the four electromagnetic form-factor constraints and satisfy the required forward-limit conditions within the reduced ansatz.  The standard dipole form is
\begin{equation}
G_D(Q^2)=\left(1+\frac{Q^2}{\Lambda_D^2}\right)^{-2},\qquad \Lambda_D^2=0.71~{\rm GeV}^2.
\label{eq:dipole}
\end{equation}
The Sachs form factors are taken as
\begin{equation}
G_E^p=G_D,\qquad G_M^p=\mu_pG_D,\qquad G_M^n=\mu_nG_D,\qquad G_E^n=-\mu_n\frac{\tau}{1+\eta\tau}G_D,\qquad \eta=5.6.
\label{eq:baseline_ff_input}
\end{equation}
These choices satisfy the correct static normalizations $G_E^p(0)=1$, $G_E^n(0)=0$, $G_M^p(0)=\mu_p$, and $G_M^n(0)=\mu_n$.  After the conversion in Eq.~\eqref{eq:sachs_to_dirac_pauli}, the flavor-separated form factors are obtained from Eq.~\eqref{eq:flavor_inverse_relation}.  These flavor form factors provide the moment constraints for $H_v^u$, $H_v^d$, $E_v^u$, and $E_v^d$.

The same reconstruction framework can be applied without modification to alternative elastic-input choices, such as Kelly-type rational parametrizations \cite{Kelly:2004hm} or $z$-expansion representations \cite{Hill:2010yb,Ye:2017gyb}.  In the present work, we focus the numerical discussion on the smooth baseline input because the purpose is to test the inverse-problem formulation and the reduced-profile MEM implementation, instead of performing a full phenomenological form-factor analysis.  A quantitative comparison among different empirical form-factor parametrizations is beyond the scope of the present proof-of-concept study and is left for future work.  A future quantitative extraction should replace the present controlled input with a global form-factor fit that incorporates experimental uncertainties, covariance information, and, where relevant, possible two-photon-exchange corrections. 

%--------------------------------------------------
\section{Forward PDF input and GPD ansatz}
\label{sec:pdf_input}
%--------------------------------------------------
The form factors alone constrain only $x$-integrated quantities.  To construct $x$-dependent GPD profiles within the present ansatz, the forward distributions must be specified.  In this baseline implementation, we use analytic beta-function parametrizations for the valence PDFs.  This choice is not meant to replace modern global PDF fits; it provides a closed analytic input that makes the structure of the MEM inverse problem transparent.  Since the present calculation is a proof-of-concept study with analytic forward inputs, all forward distributions and reconstructed GPDs are understood to be defined at a fixed reference scale $Q_0$.  QCD evolution is not included in the present baseline implementation.  In future phenomenological applications, the analytic functions can be replaced by global PDF sets such as MSTW2008, CJ, or CT-family distributions \cite{Martin:2009iq,Owens:2012bv,Dulat:2015mca}, or by more recent fits such as CT18, MSHT20, or NNPDF.

The valence PDFs are parametrized by
\begin{equation}
u_v(x)=2\frac{x^{a_u}(1-x)^{b_u}}{B(a_u+1,b_u+1)},\qquad d_v(x)=\frac{x^{a_d}(1-x)^{b_d}}{B(a_d+1,b_d+1)}.
\label{eq:pdf_beta}
\end{equation}
Here $B(\alpha,\beta)=\Gamma(\alpha)\Gamma(\beta)/\Gamma(\alpha+\beta)$.  The normalization guarantees $\int_0^1 dx\,u_v(x)=2$ and $\int_0^1 dx\,d_v(x)=1$.  The parameter $a_q$ governs the small-$x$ behavior, while $b_q$ controls the large-$x$ falloff.  The baseline values used in the present numerical test are
\begin{equation}
a_u=0.5,\qquad b_u=3.0,\qquad a_d=0.5,\qquad b_d=4.0.
\label{eq:pdf_params}
\end{equation}
The larger value of $b_d$ reflects the stronger suppression of $d_v(x)$ at large $x$, a common feature of phenomenological valence PDFs.

The forward limit of the Pauli-type GPD, $e_v^q(x)=E_v^q(x,0)$, is not an ordinary inclusive PDF.  Nevertheless, its first moment is fixed by the anomalous magnetic moment, and it must be parametrized in a normalizable form.  We use
\begin{equation}
e_v^u(x)=\kappa_u\frac{x^{a_E^u}(1-x)^{b_E^u}}{B(a_E^u+1,b_E^u+1)},\qquad e_v^d(x)=\kappa_d\frac{x^{a_E^d}(1-x)^{b_E^d}}{B(a_E^d+1,b_E^d+1)}.
\label{eq:e_beta}
\end{equation}
By construction, $\int_0^1 dx\,e_v^u(x)=\kappa_u$ and $\int_0^1 dx\,e_v^d(x)=\kappa_d$.  The baseline parameters are
\begin{equation}
a_E^u=0.5,\qquad b_E^u=4.0,\qquad a_E^d=0.5,\qquad b_E^d=5.0.
\label{eq:e_params}
\end{equation}
The additional large-$x$ suppression relative to $q_v(x)$ is a phenomenological choice that can be relaxed in future fits. Therefore, the Pauli-sector results should be interpreted together with the assumed $e^q_v(x)$ shape.

The Pauli-forward input thus introduces a model uncertainty that should be kept separate from the elastic form-factor uncertainty.  Only the first moments of $e_v^q(x)$ are fixed by the anomalous magnetic moments, while the detailed $x$-shape is parametrized.  In future quantitative fits, the parameters $a_E^q$ and $b_E^q$ should therefore be varied at fixed $\kappa_q$.  A minimal sensitivity test is, for example, $b_E^u=3.5,\ 4.0,\ 4.5$ and $b_E^d=4.5,\ 5.0,\ 5.5$ with the normalizations kept fixed.  The induced spread in $f_E^q(x)$, $E_v^q(x,t)$, and $\langle b_\perp^2\rangle_{E_q}$ then provides a quantitative estimate of the Pauli-input uncertainty.

With these forward inputs, we define the zero-skewness profile ansatz
\begin{equation}
H_v^q(x,t)=q_v(x)\exp[t f_H^q(x)],\qquad E_v^q(x,t)=e_v^q(x)\exp[t f_E^q(x)],\qquad q=u,d.
\label{eq:profile_ansatz}
\end{equation}
The ansatz automatically satisfies the forward constraints $H_v^q(x,0)=q_v(x)$ and $E_v^q(x,0)=e_v^q(x)$.  It also gives the correct normalizations of the first moments at $t=0$.  In the spacelike region, where $t=-Q^2$, this convention gives $H_v^q(x,-Q^2)=q_v(x)\exp[-Q^2 f_H^q(x)]$ and $E_v^q(x,-Q^2)=e_v^q(x)\exp[-Q^2 f_E^q(x)]$, so positive profiles generate decreasing form factors.  The corresponding form-factor sum rules are
\begin{equation}
F_1^q(t)=\int_0^1 dx\,q_v(x)\exp[t f_H^q(x)],\qquad F_2^q(t)=\int_0^1 dx\,e_v^q(x)\exp[t f_E^q(x)].
\label{eq:profile_sum_rules}
\end{equation}
The functions $f_H^q(x)$ and $f_E^q(x)$ are the unknown profile objects to be determined via the MEM.  They encode the way in which partons carrying different momentum fractions contribute to the transverse momentum-transfer dependence. % Since $t$ is measured in ${\rm GeV}^2$, the profile functions $f_H^q(x)$ and $f_E^q(x)$ have dimensions of ${\rm GeV}^{-2}$.

The expected qualitative behavior of the profile functions is physically constrained.  A large-$x$ parton tends to carry most of the nucleon longitudinal momentum and is expected to be localized closer to the transverse center of momentum.  Therefore, the transverse profile should decrease as $x\to1$. To reduce the underdetermination of the inverse problem, the final numerical implementation uses the positive three-parameter profile form
\begin{equation}
f(x)=0.05+(1-x)^2\exp(c_0+c_1x+c_2x^2).
\label{eq:reduced_profile_form}
\end{equation}
Equation~\eqref{eq:reduced_profile_form} enforces positivity, suppresses the profile at large $x$, and reduces each profile from a large set of grid values to three coefficients.  The four resulting profiles are characterized by 12 shape parameters, a reduction essential to suppressing artificial prior sensitivity in a form-factor-only inverse problem.  This profile remains finite as $x\to0$ and is not intended to encode Regge-like small-$x$ transverse growth; incorporating logarithmic or Regge-motivated terms would be a natural extension once empirical constraints at small $ x$ are included.

%--------------------------------------------------
\section{Coupled maximum-entropy reconstruction}
\label{sec:mem_formulation}
%--------------------------------------------------
The four coupled constraints are obtained by inserting the profile ansatz into the flavor decomposition.  Explicitly, the four constraints can be written compactly as
\begin{equation}
F_i^N(t_j)=\sum_{q=u,d} e_q^N\int_0^1 dx\,g_{i,v}^q(x)\,\exp\!\left[t_j f_i^q(x)\right],
\qquad i=1,2,\qquad N=p,n ,
\label{eq:common_FiN_constraint}
\end{equation}
where
\begin{equation}
g_{1,v}^q=q_v,\qquad g_{2,v}^q=e_v^q,
\qquad
f_1^q=f_H^q,\qquad f_2^q=f_E^q .
\label{eq:common_g_f_def}
\end{equation}
The charge coefficients are
\begin{equation}
(e_u^p,e_d^p)=\left(\frac{2}{3},-\frac{1}{3}\right),
\qquad
(e_u^n,e_d^n)=\left(-\frac{1}{3},\frac{2}{3}\right).
\label{eq:common_charge_coefficients}
\end{equation}
These equations show explicitly why the reconstruction should be treated as a coupled inverse problem.  Proton and neutron form factors mix $u$- and $d$-flavor contributions with different charge weights, while the Dirac and Pauli sectors constrain different GPD structures.

The MEM functional is defined as
\begin{equation}
Q=\alpha S-\frac{1}{2}\chi^2.
\label{eq:mem_functional}
\end{equation}
The likelihood term quantifies the deviation between the input form factors and those reconstructed from the GPDs.  If the full covariance matrix is available, one may write
\begin{equation}
\chi^2=\sum_{a,b}\sum_{i,j}\left[F_a^{\rm input}(t_i)-F_a^{\rm MEM}(t_i)\right]C^{-1}_{ai,bj}\left[F_b^{\rm input}(t_j)-F_b^{\rm MEM}(t_j)\right],
\label{eq:chi_cov}
\end{equation}
where $a,b$ label the four form-factor channels.  In the present controlled calculation, where a smooth elastic input is used, we employ the diagonal approximation
\begin{equation}
\chi^2=\sum_a\sum_i\frac{\left[F_a^{\rm input}(t_i)-F_a^{\rm MEM}(t_i)\right]^2}{\sigma_a^2(t_i)}.
\label{eq:chi_diag}
\end{equation}
The assigned uncertainties govern the numerical strength of the form-factor constraints and enable us to assess the stability of the inverse reconstruction.  For the controlled smooth form-factor input used here, these quantities are assigned as numerical tolerances,
\begin{equation}
\sigma_a(t_i)
=
\epsilon\,\max\left(|F_a^{\rm input}(t_i)|,F_{\rm floor}\right),
\qquad
 a=F_1^p,F_1^n,F_2^p,F_2^n .
\label{eq:sigma_assignment}
\end{equation}
In the numerical calculation, we use
\begin{equation}
\epsilon=0.03,
\qquad
F_{\rm floor}=0.02 .
\label{eq:sigma_numerical_values}
\end{equation}
These assigned uncertainties should not be interpreted as experimental errors.  They only specify the numerical strength with which the smooth input form factors are imposed in the present proof-of-concept reconstruction.

For positive profile functions, the Shannon-Jaynes entropy is
\begin{equation}
S[f,m]=\int_0^1 dx\,\left[f(x)-m(x)-f(x)\ln\frac{f(x)}{m(x)}\right].
\label{eq:entropy}
\end{equation}
The total entropy is
\begin{equation}
S=\sum_{q=u,d}\big[S[f_H^q,m_H^q]+S[f_E^q,m_E^q]\big].
\label{eq:total_entropy}
\end{equation}
The entropy term penalizes unnecessary deviations from the default profile while allowing deviations required by the form-factor constraints.  The hyperparameter $\alpha$ controls the balance between data fidelity and prior smoothness.  The default model used in the numerical calculation is a positive, large-$x$-suppressed profile,
\begin{equation}
m_A(x)=a_A(1-x)^2+0.05,
\qquad
A=H^u,H^d,E^u,E^d .
\label{eq:default_profile_model}
\end{equation}
The channel-dependent amplitudes are chosen as
\begin{equation}
a_{H^u}=1.0,
\qquad
a_{H^d}=1.1,
\qquad
a_{E^u}=1.2,
\qquad
a_{E^d}=1.3 .
\label{eq:default_profile_amplitudes}
\end{equation}
The default model in Eq.~\eqref{eq:default_profile_model} serves only as a smooth regularizing prior in the entropy functional and is not imposed as the final transverse profile.

The entropy functional is applied to the positive profile functions $f_H^q(x)$ and $f_E^q(x)$ themselves, not to the signed GPDs $H_v^q(x,t)$ and $E_v^q(x,t)$, i.e., a distinction that matters for the $d$-quark Pauli sector.  Since $\kappa_d<0$, the forward function $e_v^d(x)$ and the corresponding $E_v^d(x,t)$ are negative in the present convention.  The entropy-regularized quantity, however, is the positive transverse profile $f_E^d(x)$.  The Shannon-Jaynes entropy is therefore used consistently, while $E_v^d$ is interpreted as a signed transverse distribution.

A grid-based reconstruction of $f_H^q(x)$ and $f_E^q(x)$ has many local degrees of freedom and can become prior sensitive, because elastic form factors constrain only $x$-integrated moments.  To stabilize the inverse problem, the numerical implementation adopts the reduced profile form in Eq.~\eqref{eq:reduced_profile_form}: the unknowns become the three coefficients $(c_0,\,c_1,\,c_2)$ for each of the four profiles rather than independent values at every grid point, which preserves positivity and large-$x$ suppression while removing unconstrained local oscillations.  The entropy term then regularizes this physically constrained low-dimensional manifold relative to the default model, so the present calculation is an entropy-regularized profile reconstruction on a reduced positive manifold.  In this implementation, the form-factor constraints determine the fitted profile coefficients, while the entropy functional preserves positivity and provides a controlled default-model regularization.

The profile coefficients are determined by maximizing $Q$ over the twelve-dimensional parameter space.  The resulting profiles define the reconstructed GPDs,
\begin{equation}
H_{v,{\rm MEM}}^q(x,t)=q_v(x)\exp[t f_{H,{\rm MEM}}^q(x)],\qquad E_{v,{\rm MEM}}^q(x,t)=e_v^q(x)\exp[t f_{E,{\rm MEM}}^q(x)].
\label{eq:reconstructed_gpd}
\end{equation}
Hence, the reconstruction should be interpreted as the least-biased profile solution compatible with the elastic moment constraints, the forward PDF input, the anomalous magnetic moment normalizations, and the reduced positive profile ansatz.

%--------------------------------------------------
\section{Discretized numerical scheme}
\label{sec:numerical_scheme}
%--------------------------------------------------
For numerical implementation, the interval $0\le x\le1$ is discretized as $x_j$ with quadrature weights $w_j$, while the spacelike interval $0\le Q^2\le5~{\rm GeV}^2$ is discretized into uniform points.  In the numerical results shown below, we use 100 intervals in both variables, corresponding to 101 points in $x$ and 101 points in $Q^2$.  The continuous integrals are replaced by finite sums.  To avoid a spurious mismatch at $Q^2=0$ from finite-grid quadrature, the forward inputs are normalized on the same discrete grid used in the reconstruction:
\begin{equation}
\sum_j w_j u_v(x_j)=2,\qquad
\sum_j w_j d_v(x_j)=1,\qquad
\sum_j w_j e_v^u(x_j)=\kappa_u,\qquad
\sum_j w_j e_v^d(x_j)=\kappa_d .
\label{eq:discrete_forward_normalization}
\end{equation}
Imposing this discrete normalization in advance ensures that the static Sachs and flavor-form-factor normalizations are exact to machine precision before the nonzero-$Q^2$ profile fit is performed.  For example,
\begin{equation}
F_1^{u,{\rm MEM}}(t_i)=\sum_{j=1}^{N_x}w_j u_v(x_j)\exp[t_i f_H^u(x_j)].
\label{eq:discrete_F1u}
\end{equation}

The profile functions are not treated as independent values at all $x_j$.  Instead, for each of $f_H^u$, $f_H^d$, $f_E^u$, and $f_E^d$, we use
\begin{equation}
f_A(x;c_0^A,c_1^A,c_2^A)=0.05+(1-x)^2\exp(c_0^A+c_1^Ax+c_2^Ax^2),
\qquad A=H^u,H^d,E^u,E^d.
\label{eq:discrete_reduced_profile}
\end{equation}
Relative to a direct grid representation, which would contain $4N_x$ local profile values, this is a substantial reduction; it matters because form factors provide moment constraints rather than direct local measurements in $x$.

The entropy functional is evaluated with respect to the default model $m_A(x)$ in Eq.~\eqref{eq:default_profile_model}.  The initial profile is chosen to be close to this default model.  In practice, the initial parameters $c_0^A,c_1^A,c_2^A$ are selected by minimizing
\begin{equation}
\int_0^1 dx\,
\left[
 f_A(x;c_0^A,c_1^A,c_2^A)-m_A(x)
\right]^2
\label{eq:initial_profile_matching}
\end{equation}
for each channel $A$.  The result is a smooth and positive initial condition, consistent with the large-$x$-suppressed default profile, that remains inside the reduced profile manifold.

The entropy derivative has the simple form
\begin{equation}
\frac{\delta S[f,m]}{\delta f(x)}=-\ln\frac{f(x)}{m(x)}.
\label{eq:entropy_derivative}
\end{equation}
The form-factor gradients with respect to the profile values are analytic, for instance
\begin{equation}
\frac{\partial F_1^{u,{\rm MEM}}(t_i)}{\partial f_H^u(x_j)}=w_j u_v(x_j)t_i\exp[t_i f_H^u(x_j)].
\label{eq:gradient_F1u}
\end{equation}
The gradients with respect to the reduced coefficients follow by the chain rule, and these analytic expressions keep the optimization stable while avoiding numerical finite-difference noise.  Entropy maximization is achieved by minimizing $-Q$ using the L-BFGS-B algorithm.  %The stopping parameters are\begin{equation}{\tt maxiter}=120,\qquad{\tt ftol}=10^{-10},\qquad{\tt gtol}=10^{-7}.\label{eq:optimizer_settings}\end{equation}
The hyperparameter $\alpha$ is varied around the reference value $\alpha_0=10^{-3}$.  Stable features under this variation are interpreted as constraint-driven, while unstable features would indicate residual dependence on the entropy prior.  The reduced profile representation is designed to minimize such artificial prior dependence by removing local profile modes that are not constrained by the elastic moments.

%--------------------------------------------------
\section{Physical interpretation and consistency checks}
\label{sec:interpretation}
%--------------------------------------------------
The main physical output of the reconstruction is the set of transverse profiles $f_H^q(x)$ and $f_E^q(x)$.  At zero skewness, $H_v^q(x,t)$ can be Fourier transformed to impact-parameter space:
\begin{equation}
q_v(x,\bm b_\perp)=\int\frac{d^2\bm\Delta_\perp}{(2\pi)^2}\exp[-i\bm\Delta_\perp\cdot\bm b_\perp]H_v^q(x,0,-\bm\Delta_\perp^2).
\label{eq:impact_parameter}
\end{equation}
For the exponential ansatz, the small-$|t|$ slope gives
\begin{equation}
\langle b_\perp^2\rangle_x^q\simeq4\left.\frac{\partial}{\partial t}\ln H_v^q(x,t)\right|_{t=0}=4f_H^q(x).
\label{eq:b2}
\end{equation}
So $f_H^q(x)$ has a direct interpretation as the $x$-dependent transverse size of the valence distribution.  A decreasing $f_H^q(x)$ with increasing $x$ would support the physical picture that large-$x$ quarks are more localized near the transverse center of momentum.

The Pauli-type profile $f_E^q(x)$ has a different interpretation because $E_v^q$ contributes to the spin-flip structure.  In a transversely polarized nucleon, $E$ is associated with transverse distortions of the parton distribution and is related to orbital angular momentum effects through the broader GPD framework \cite{Ji:1996ek,Burkardt:2000za,Diehl:2003ny}.  Although the present work does not attempt a full angular-momentum decomposition, a stable extraction of $E_v^q(x,t)$ would provide useful input for such analyses.

Several consistency checks are required.  First, the reconstructed GPDs must reproduce the input form factors:
\begin{equation}
\int_0^1 dx\,H_{v,{\rm MEM}}^{q,X}(x,t)=F_{1,X}^q(t),\qquad \int_0^1 dx\,E_{v,{\rm MEM}}^{q,X}(x,t)=F_{2,X}^q(t),
\label{eq:ff_check}
\end{equation}
within the assigned uncertainties.  Second, the forward-limit PDF check must hold:
\begin{equation}
q_{v,{\rm rec}}^q(x)\equiv H_{v,{\rm MEM}}^q(x,0)=q_v(x).
\label{eq:pdf_check}
\end{equation}
In the present ansatz, Eq.~\eqref{eq:pdf_check} is imposed exactly.  In a more general reconstruction, where the forward PDF is allowed to fluctuate within a prior uncertainty band, Eq.~\eqref{eq:pdf_check} becomes a nontrivial posterior test.  Deviations may be quantified by $\Delta q_v^q(x)=H_{v,{\rm MEM}}^q(x,0)-q_v^{\rm input}(x)$ or $R_q(x)=H_{v,{\rm MEM}}^q(x,0)/q_v^{\rm input}(x)$.

Third, the dependence on the elastic-input parametrization should be assessed in a quantitative phenomenological application.  The present study uses the baseline input of Eq.~\eqref{eq:baseline_ff_input}, while the formalism is directly compatible with Kelly-type or $z$-expansion form-factor inputs.  In a future extraction, this check should be performed using published covariance information and reported as a separate elastic-input systematic uncertainty.

%--------------------------------------------------
\section{Numerical results}
\label{sec:numerical_results}
%--------------------------------------------------
In this section, we present the numerical reconstruction obtained from the baseline elastic input and the reduced positive profile ansatz.  We first test the stability of the MEM solution, then check the forward normalization and electromagnetic form-factor moments.  We next discuss the impact-parameter-space distributions and the associated transverse mean-square radii.  Finally, we present a qualitative proxy for beam-spin asymmetry and summarize the main systematic uncertainties in the present reconstruction.

%-----------------------------------------------------------
\subsection{Stability under variations of the entropy weight}
\label{subsec:alpha_stability}
%-----------------------------------------------------------
The reduced-profile MEM fit contains the hyperparameter $\alpha$, which controls the balance between the entropy prior and the form-factor likelihood.  To test the stability of the reconstructed profiles, we vary $\alpha$ around the reference value $\alpha_0=10^{-3}$ and compare the resulting profiles with the reference solution.  For each profile $f(x;\alpha)$, we define
\begin{equation}
\delta_\alpha(x)=100\times\left[\frac{f(x;\alpha)}{f(x;\alpha_0)}-1\right],
\qquad \alpha_0=10^{-3},
\label{eq:alpha_pointwise_deviation}
\end{equation}
and the integrated relative deviation
\begin{equation}
\Delta_\alpha[f]=
\left[
\frac{\int_0^1 dx\,\left[f(x;\alpha)-f(x;\alpha_0)\right]^2}
{\int_0^1 dx\,\left[f(x;\alpha_0)\right]^2}
\right]^{1/2}.
\label{eq:alpha_integrated_deviation}
\end{equation}
The maximum pointwise deviation is also evaluated over the full $x$-range and over the central region $0.1\le x\le0.9$.
\begin{table}[b]
\begin{tabular}{ccccccccc}
\hline\hline
Profile &
$c_0$ & $c_1$ & $c_2$ &
\multicolumn{2}{c}{$100\Delta_\alpha$ [\%]} &
\multicolumn{2}{c}{$\max_x|\delta_\alpha|$ [\%]} &
$\max_{0.1\le x\le0.9}|\delta_\alpha|$ [\%] \\
\cline{5-8}
 & & & &
$\alpha=10^{-4}$ & $\alpha=10^{-2}$ &
$\alpha=10^{-4}$ & $\alpha=10^{-2}$ &
$\alpha=10^{-2}$ \\
\hline
$f_H^u$ & 1.673 & -1.960 & 0.6464 &
0.0000 & 0.0000 & 0.0000 & 0.0010 & 0.0010 \\
$f_H^d$ & 1.931 & -5.220 & 7.796 &
0.0000 & 0.0040 & 0.0010 & 0.0060 & 0.0060 \\
$f_E^u$ & 1.979 & -2.590 & 2.022 &
0.0000 & 0.0000 & 0.0000 & 0.0010 & 0.0010 \\
$f_E^d$ & 1.657 & -2.891 & 1.850 &
0.0000 & 0.0000 & 0.0000 & 0.0000 & 0.0000 \\
\hline\hline
\end{tabular}
\caption{Reduced-profile coefficients and $\alpha$-stability of the reconstructed zero-skewness valence GPD profiles.  The reference profile is obtained at $\alpha_0=10^{-3}$ with $f_A(x)=0.05+(1-x)^2\exp(c_0^A+c_1^A x+c_2^A x^2)$, where $A=H^u,H^d,E^u,E^d$.  The quantities $100\Delta_\alpha$ and $\max|\delta_\alpha|$ denote the integrated and maximum pointwise percentage deviations from the reference profile, respectively.}
\label{tab:alpha_stability_reduced_profile}
\end{table}

The reduced three-parameter profiles are highly stable under the tested variation of $\alpha$.  In particular, Table~\ref{tab:alpha_stability_reduced_profile} shows that the maximum pointwise deviation remains below $0.01\%$ for all channels.  The small $\alpha$-dependence reflects the fact that the reduced profile manifold removes most local modes that are not constrained by elastic moments.  Within this manifold, the form-factor likelihood determines the fitted coefficients, while the entropy term preserves positivity and provides default-model regularization.  In less restricted profile representations or with noisier empirical inputs, the same entropy functional would play a more prominent role in suppressing unconstrained local structures.

%-----------------------------------------------------------
\subsection{Forward normalization and form-factor consistency}
\label{subsec:ff_consistency}
%-----------------------------------------------------------
The reconstructed GPDs must satisfy the forward normalizations and reproduce the input electromagnetic form factors through their first moments.  At $\xi=0$ and $t=0$, the relevant sum rules are
\begin{equation}
\int_0^1 dx\,\left[H_v^u(x,0,0),\,H_v^d(x,0,0),\,E_v^u(x,0,0),\,E_v^d(x,0,0)\right]
=(2,\,1,\,\kappa_u,\,\kappa_d).
\label{eq:forward_normalization_check}
\end{equation}
With the discrete-grid normalization of Eq.~\eqref{eq:discrete_forward_normalization}, the corresponding numerical values are $2.000$, $1.000$, $1.673$, and $-2.033$, respectively, up to the printed precision.  The numerical GPDs satisfy the required valence-number and Pauli-form-factor normalization conditions exactly at the grid level.  The remaining nonzero-$Q^2$ differences in Table~\ref{tab:input_vs_gpd_form_factors} therefore reflect the restricted profile fit, not a failure of the forward normalization.

\begin{figure}[t]
\topinset{(a)}{\includegraphics[width=4.0cm]{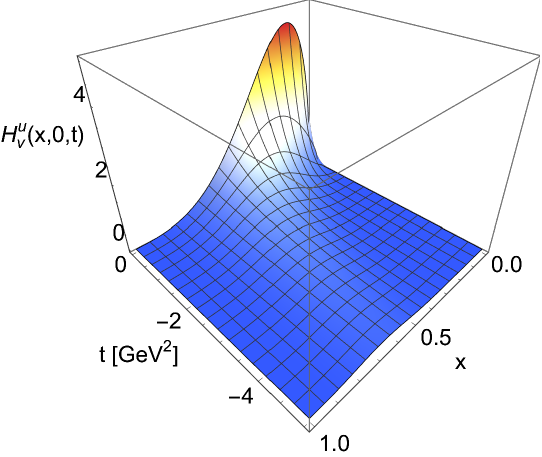}}{-0.4cm}{0.0cm}
\topinset{(b)}{\includegraphics[width=4.0cm]{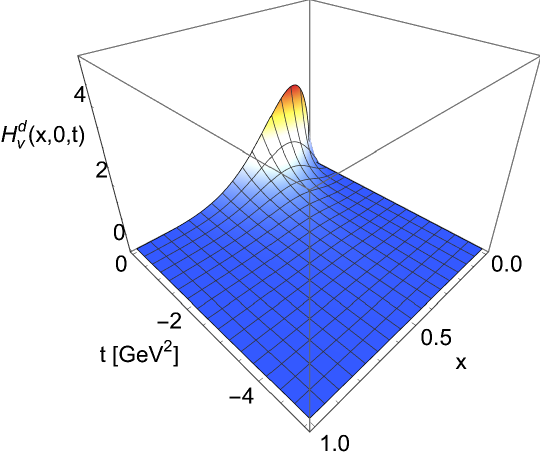}}{-0.4cm}{0.0cm}
\topinset{(c)}{\includegraphics[width=4.0cm]{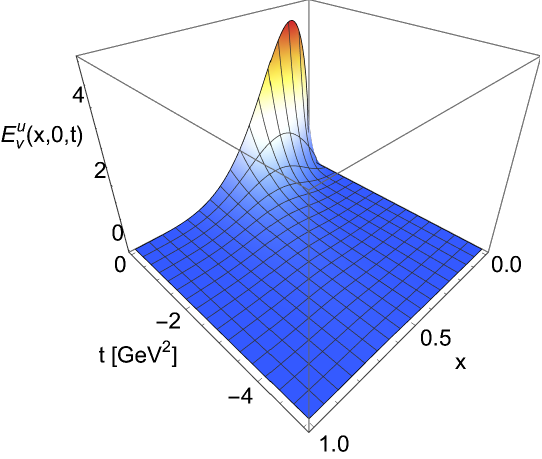}}{-0.4cm}{0.0cm}
\topinset{(d)}{\includegraphics[width=4.0cm]{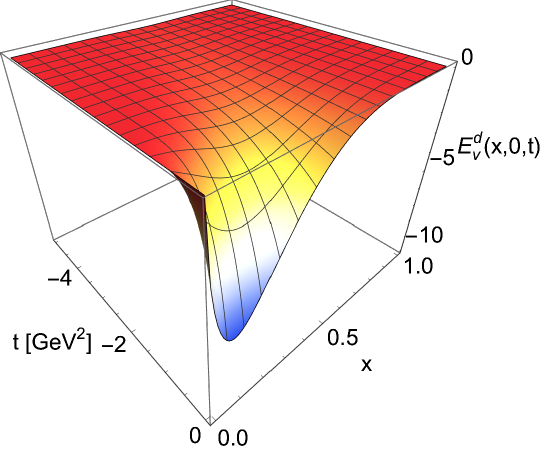}}{-0.4cm}{0.0cm}
\\
\topinset{(e)}{\includegraphics[width=4.0cm]{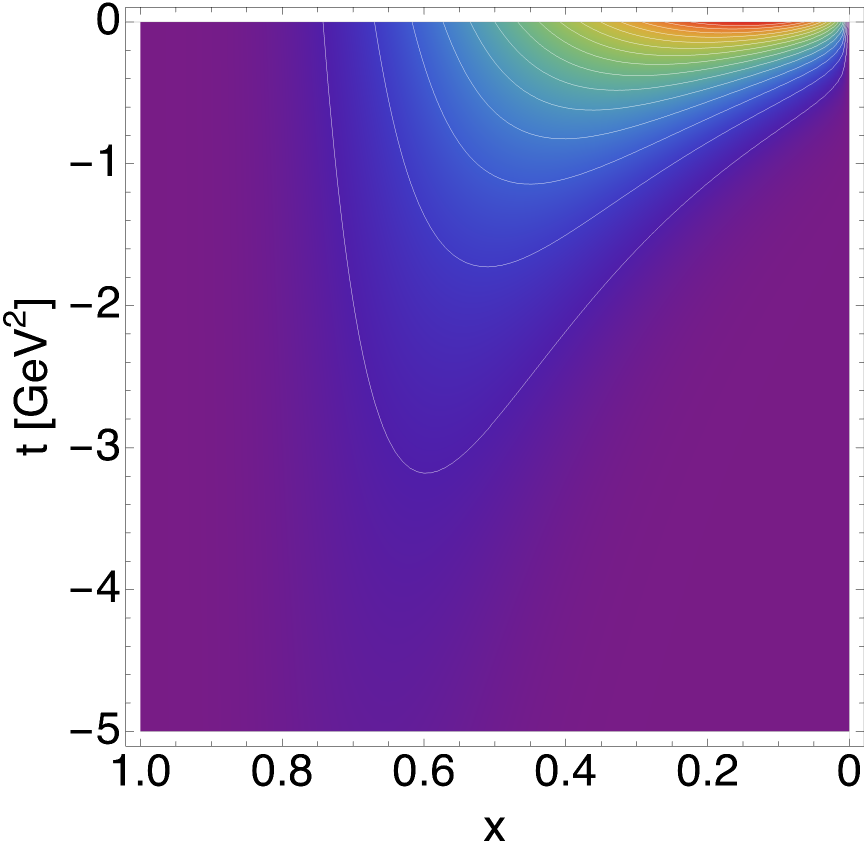}}{-0.4cm}{0.0cm}
\topinset{(f)}{\includegraphics[width=4.0cm]{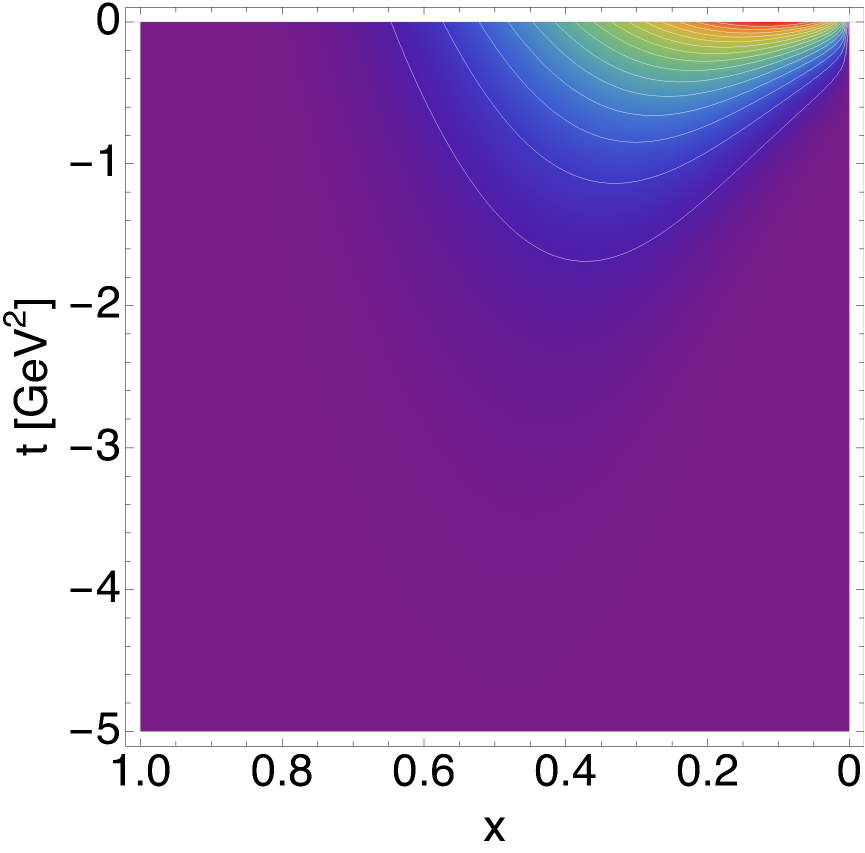}}{-0.4cm}{0.0cm}
\topinset{(g)}{\includegraphics[width=4.0cm]{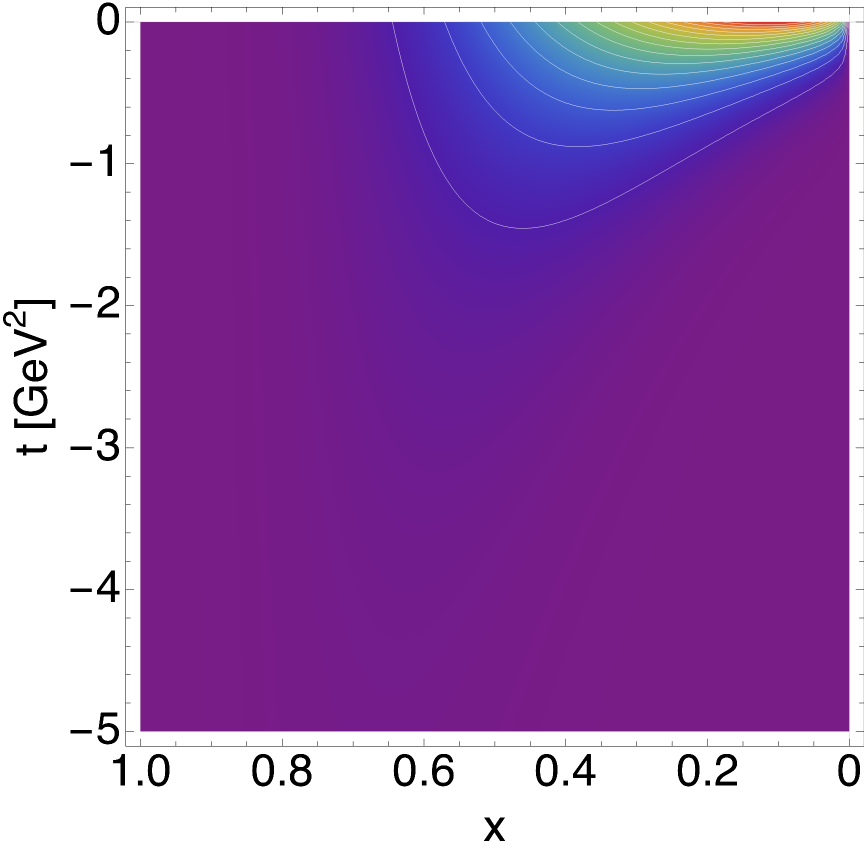}}{-0.4cm}{0.0cm}
\topinset{(h)}{\includegraphics[width=4.0cm]{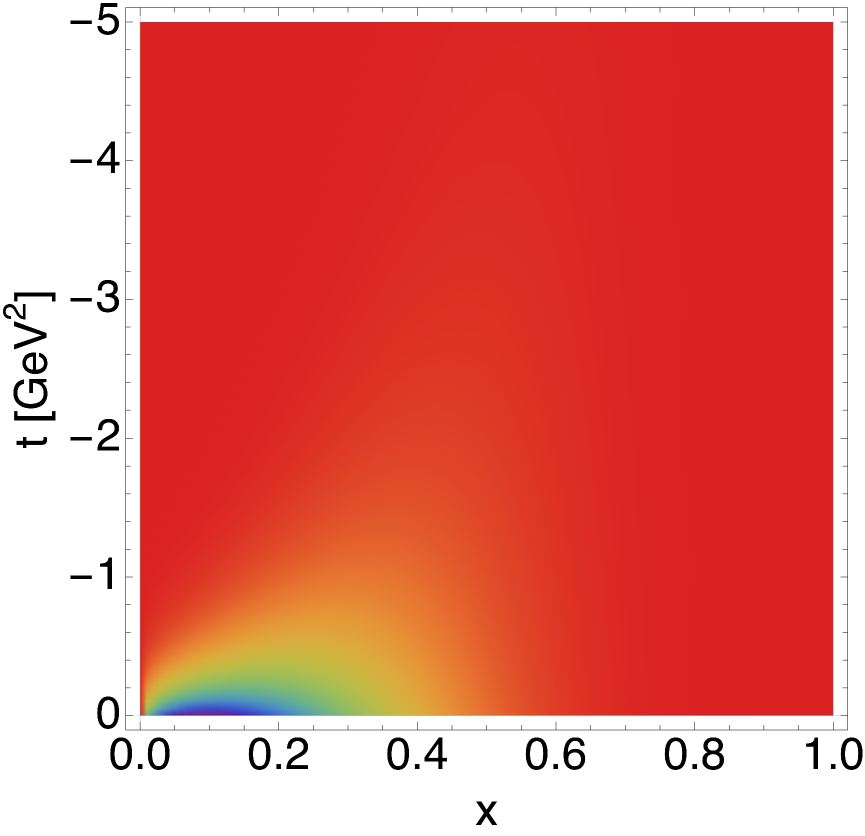}}{-0.4cm}{0.0cm}
\caption{(Color online) Reconstructed zero-skewness valence GPDs constructed from the reduced-profile MEM solution using the forward PDF input and electromagnetic form-factor constraints: (a) $H^u_v(x,0,t)$, (b) $H^d_v(x,0,t)$, (c) $E^u_v(x,0,t)$, and (d) $E^d_v(x,0,t)$.  Panels (e)-(h) show the corresponding two-dimensional maps.}
\label{FIG1}
\end{figure}

Fig.~\ref{FIG1} shows the reconstructed zero-skewness valence GPDs generated by the reduced-profile MEM solution.
The upper panels display the three-dimensional $x$- and $t$-dependence, while the lower panels show the corresponding two-dimensional maps. The Dirac-type distributions reduce to the input valence PDFs at $t=0$, whereas the Pauli-type distributions reduce to $e_v^q(x)$ with anomalous-moment normalization. The suppression as $Q^2=-t$ increases reflects the positive transverse profiles in the exponential ansatz. As noted in Sec.~\ref{sec:mem_formulation}, the $E_v^d$ surface is negative simply because $\kappa_d<0$ fixes its forward normalization.

\begin{table}[b]
\begin{tabular}{c|cc|cc|cc|cc}
\hline\hline
$Q^2$ [${\rm GeV}^2$] &
$G_E^p$ input & $G_E^p$ GPD &
$G_M^p$ input & $G_M^p$ GPD &
$G_E^n$ input & $G_E^n$ GPD &
$G_M^n$ input & $G_M^n$ GPD \\
\hline
0.00 & 1.000 & 1.000 & 2.793 & 2.793 & 0.000 & 0.000 & $-1.913$ & $-1.913$ \\
0.50 & 0.344 & 0.348 & 0.962 & 0.968 & 0.052 & 0.048 & $-0.659$ & $-0.667$ \\
1.00 & 0.172 & 0.171 & 0.481 & 0.478 & 0.036 & 0.038 & $-0.330$ & $-0.327$ \\
2.00 & 0.069 & 0.068 & 0.192 & 0.191 & 0.018 & 0.018 & $-0.131$ & $-0.131$ \\
3.00 & 0.037 & 0.037 & 0.102 & 0.103 & 0.010 & 0.010 & $-0.070$ & $-0.071$ \\
4.00 & 0.023 & 0.023 & 0.063 & 0.063 & 0.007 & 0.007 & $-0.043$ & $-0.043$ \\
5.00 & 0.015 & 0.015 & 0.043 & 0.042 & 0.005 & 0.005 & $-0.030$ & $-0.029$ \\
\hline\hline
\end{tabular}
\caption{Comparison between the input Sachs form factors and the values reconstructed from the GPD moments at selected $Q^2$ points.}
\label{tab:input_vs_gpd_form_factors}
\end{table}

The agreement between the input and reconstructed form factors is at the percent level for $G_E^p$, $G_M^p$, and $G_M^n$.  The relative difference for $G_E^n$ can appear larger because $G_E^n$ is numerically small, but the absolute difference remains small over the tested range; for $G_E^n$, the absolute residual is the more meaningful diagnostic precisely because the form factor itself crosses a numerically small range.

%-----------------------------------------------------------
\subsection{Impact-parameter-space distributions}
\label{subsec:impact_parameter_results}
%-----------------------------------------------------------
Since the reconstructed GPDs are obtained at zero skewness, they can be transformed into impact-parameter space.  For the Dirac-type valence GPD, we define
\begin{equation}
q_v(x,\mathbf b_\perp)=
\int \frac{d^2\boldsymbol{\Delta}_\perp}{(2\pi)^2}
e^{-i\boldsymbol{\Delta}_\perp\cdot \mathbf b_\perp}
H_v^q(x,0,-\boldsymbol{\Delta}_\perp^2),
\label{eq:impact_transform_H}
\end{equation}
where $\mathbf b_\perp=(b_x,b_y)$ and $b^2_\perp=b_x^2+b_y^2$.  For the exponential profile used in the
reconstruction, $H_v^q(x,0,t)=q_v(x)\exp[t f_H^q(x)]$, with $t=-\boldsymbol{\Delta}_\perp^2$, the Fourier transform can be performed analytically and gives
\begin{equation}
q_v(x,\mathbf b_\perp)=\frac{q_v(x)}{4\pi f_H^q(x)}
\exp\left[
-\frac{b_x^2+b_y^2}{4f_H^q(x)}
\right].
\label{eq:impact_density_H}
\end{equation}
The corresponding Pauli-type distribution is obtained analogously as
\begin{equation}
e_v^q(x,\mathbf b_\perp)=\frac{e_v^q(x)}{4\pi f_E^q(x)}
\exp\left[
-\frac{b_x^2+b_y^2}{4f_E^q(x)}
\right].
\label{eq:impact_density_E}
\end{equation}
In Eqs.~(\ref{eq:impact_density_H}) and (\ref{eq:impact_density_E}), $b_\perp$ is expressed in natural units.  In the figures and in Table~\ref{tab:bperp2_integrated}, the corresponding lengths are converted to fm using $1~{\rm GeV}^{-1}=0.1973~{\rm fm}$.  Thus, the profile functions $f_H^q(x)$ and $f_E^q(x)$ determine the transverse spatial widths of the reconstructed distributions.  In particular,
\begin{equation}
\langle b_\perp^2\rangle_x^q=4f_H^q(x)
\label{eq:bperp2_x_relation}
\end{equation}
for the Dirac sector, with an analogous relation for the Pauli sector. 

\begin{figure}[!t]
\includegraphics[width=18cm]{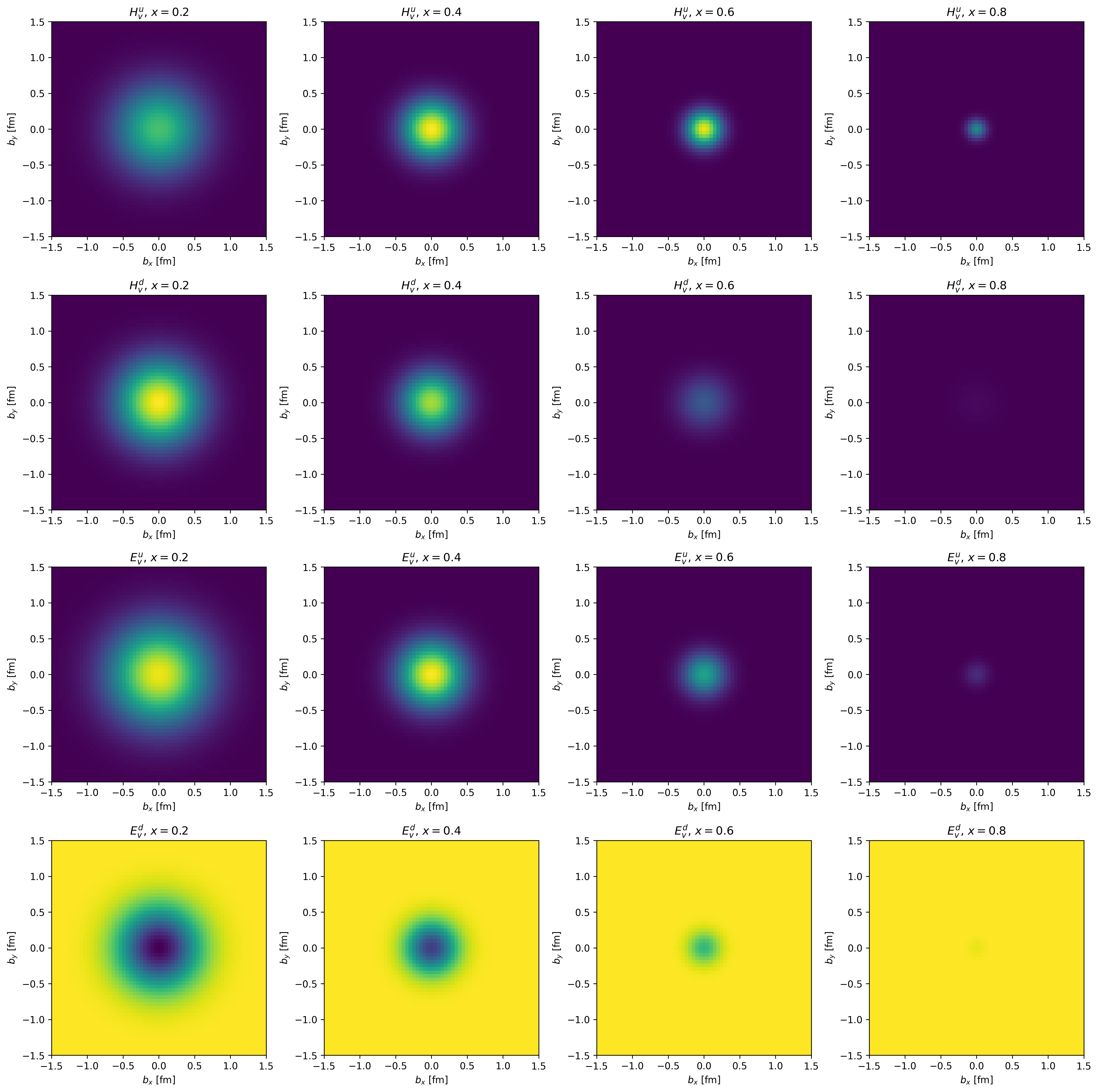}
\caption{(Color online) Impact-parameter-space distributions of the reconstructed valence GPDs in the transverse $b_x$-$b_y$ plane. The rows show, from top to bottom, $H_v^u$, $H_v^d$, $E_v^u$, and $E_v^d$, while the columns correspond to $x=0.2,0.4,0.6$, and $0.8$. The Dirac-type distributions $H_v^u$ and $H_v^d$ represent transverse valence-quark densities at fixed longitudinal momentum fraction $x$. The Pauli-type distributions are associated with the
anomalous magnetic structure; in particular, $E_v^d$ is a signed negative distribution in the present convention and should not be interpreted as a positive-definite density. As $x$ increases, the distributions become increasingly localized near $b_\perp=0$, showing the expected transverse shrinkage of large-$x$ partons.}
\label{FIG3}
\end{figure}

Fig.~\ref{FIG3} shows the impact-parameter-space distributions obtained from the reconstructed zero-skewness GPDs.
The four rows correspond to $H_v^u$, $H_v^d$, $E_v^u$, and $E_v^d$, and the four columns show representative longitudinal momentum fractions, $x=0.2,0.4,0.6$, and $0.8$. For the Dirac-type channels, the distributions can be interpreted as transverse valence-quark densities at fixed $x$. The narrowing of the profiles with increasing $x$ is clearly visible and reflects the decrease of the transverse profile functions toward the large-$x$ region, consistent with the physical expectation that a parton carrying a larger fraction of the nucleon longitudinal momentum is localized closer to the transverse center of momentum. For the Pauli-type channels, the distributions describe the transverse structure associated with the spin-flip GPD $E_v^q$. 

\begin{table}[b]
\begin{tabular}{cccc}
\hline\hline
Channel &
$\langle b_\perp^2\rangle$ [GeV$^{-2}$] &
$\langle b_\perp^2\rangle$ [fm$^{2}$] &
$\sqrt{\langle b_\perp^2\rangle}$ [fm] \\
\hline
$H_v^u$ & 8.738411 & 0.340276 & 0.583332 \\
$H_v^d$ & 10.178608 & 0.395994 & 0.629281 \\
$E_v^u$ & 12.801044 & 0.498602 & 0.706117 \\
$E_v^d$ & 9.806013 & 0.381687 & 0.617810 \\
\hline\hline
\end{tabular}
\caption{$x$-integrated transverse mean-square radii obtained from the reconstructed zero-skewness GPD profiles.  The averages are computed as $\langle b_\perp^2\rangle^q=\int_0^1 dx\,w^q(x)\,4f^q(x)/\int_0^1 dx\,w^q(x)$, where $w^q=q_v$ for the Dirac sector and $w^q=e_v^q$ for the Pauli sector.  For $E_v^d$, the average is evaluated with the signed weight $e_v^d(x)$; the positive value reflects the transverse width of the signed Pauli distribution.}
\label{tab:bperp2_integrated}
\end{table}

%-----------------------------------------------------------
\subsection{Beam-spin-asymmetry proxy as a qualitative check}
\label{subsec:dvcs_proxy}
%-----------------------------------------------------------
As a qualitative check beyond the elastic form-factor moments, we compare the reconstructed profiles with the leading azimuthal pattern of the beam-spin asymmetry in exclusive photon electroproduction,
\begin{equation}
e p \to e' p' \gamma .
\end{equation}
The beam-spin asymmetry is defined by
\begin{equation}
A_{LU}(\phi)=
\frac{d\sigma^{+}(\phi)-d\sigma^{-}(\phi)}
{d\sigma^{+}(\phi)+d\sigma^{-}(\phi)} ,
\end{equation}
where $+$ and $-$ denote the electron beam helicity states, and $\phi$ is the azimuthal angle between the lepton scattering plane and the hadronic production plane.  In the DVCS kinematic region, the measured amplitude receives contributions from the Bethe-Heitler and DVCS processes, as well as their interference.  The beam-helicity-dependent part of the cross section is dominated, at leading twist, by the imaginary part of the Bethe-Heitler--DVCS interference term \cite{Vanderhaeghen:1999xj,Guidal:2013rya,Kumericki:2016ehc}.  We use the CLAS12 beam-spin-asymmetry data of Ref.~\cite{CLAS:2022syx} only to check the leading $\sin\phi$-type behavior; these data are not included as constraints in the MEM reconstruction.

The constructed profiles are defined at zero skewness, $H_v^q(x,0,t)$ and $E_v^q(x,0,t)$.  For this diagnostic comparison, we use the minimal identification
\begin{equation}
\xi \simeq \frac{x_B}{2-x_B},
\qquad
H_v^q(\xi,\xi,t)\simeq H_v^q(\xi,0,t),
\qquad
E_v^q(\xi,\xi,t)\simeq E_v^q(\xi,0,t).
\end{equation}
Restricting to this minimal identification keeps only the information available in the present zero-skewness construction.  The imaginary parts of the dominant CFFs are then estimated as
\begin{eqnarray}
{\rm Im}\,\mathcal H(\xi,t)
&\simeq&
-\pi\left[
\frac{4}{9}H_v^u(\xi,0,t)+\frac{1}{9}H_v^d(\xi,0,t)
\right],
\\
{\rm Im}\,\mathcal E(\xi,t)
&\simeq&
-\pi\left[
\frac{4}{9}E_v^u(\xi,0,t)+\frac{1}{9}E_v^d(\xi,0,t)
\right].
\end{eqnarray}
The overall sign follows the CFF convention adopted in the present diagnostic definition.  Since the normalization factor $K$ is fitted below, the proxy is not used to infer an independent sign convention for the physical DVCS amplitude.
Retaining only the unpolarized $H$ and $E$ contributions and omitting the $\widetilde H$ term in this diagnostic proxy, the unpolarized interference combination is then approximated by
\begin{equation}
{\rm Im}\,\mathcal C_{\rm unp}^{I}\simeq 
F_1^p(t)\,{\rm Im}\,\mathcal H-\frac{t}{4M_N^2}F_2^p(t)\,{\rm Im}\,\mathcal E .
\end{equation}
We use the one-parameter proxy
\begin{equation}
A_{LU}^{\rm proxy}(\phi)=K\,{\rm Im}\,\mathcal C_{\rm unp}^{I}\sin\phi .
\label{eq:ALU}
\end{equation}
The parameter $K$ fixes the overall scale and is determined by a weighted least-squares fit,
\begin{equation}
\chi^2(K)=\sum_i\left[\frac{A_{LU,i}^{\rm exp}
-K\,{\rm Im}\mathcal C_{{\rm unp},i}^{I}\sin\phi_i}{\sigma_i}\right]^2 ,
\end{equation}
which gives
\begin{equation}
K=\frac{\sum_i A_{LU,i}^{\rm exp}\,{\rm Im}\mathcal C_{{\rm unp},i}^{I}\sin\phi_i/\sigma_i^2}
{\sum_i\left[{\rm Im}\mathcal C_{{\rm unp},i}^{I}\sin\phi_i\right]^2/\sigma_i^2}.
\end{equation}
The fitted constant $K$ sets the proxy's overall scale in this simplified comparison.

\begin{table}[b]
\begin{tabular}{lccc}
\hline\hline
Data set & $K$ & Number of points & $\chi^2/{\rm dof}$ \\
\hline
All data & $-0.136$ & 1535 & 2.735 \\
$E_b=10.2~{\rm GeV}$ & $-0.141$ & 529 & 3.043 \\
$E_b=10.6~{\rm GeV}$ & $-0.134$ & 1006 & 2.573 \\
\hline\hline
\end{tabular}
\caption{Weighted least-squares fit results for the beam-spin-asymmetry proxy $A_{LU}^{\rm proxy}(\phi)=K\,{\rm Im}\,\mathcal C_{\rm unp}^{I}\sin\phi$.  The fitted constant $K$ sets the overall scale of this qualitative comparison.}
\label{tab:dvcs_proxy_fit}
\end{table}

The fitted normalization factors are summarized in Table~\ref{tab:dvcs_proxy_fit}.  The full data set gives $K=-0.136$ with $\chi^2/{\rm dof}=2.735$.  Separate fits to the $E_b=10.2~{\rm GeV}$ and $E_b=10.6~{\rm GeV}$ subsets give $K=-0.141$ and $K=-0.134$, respectively.  The similarity among these three values indicates that the proxy's overall scale is stable across this beam-energy separation, while the $\chi^2/{\rm dof}$ values quantify the limited scope of this one-parameter comparison.

\begin{figure}[t]
\topinset{(a)}{\includegraphics[width=7.5cm]{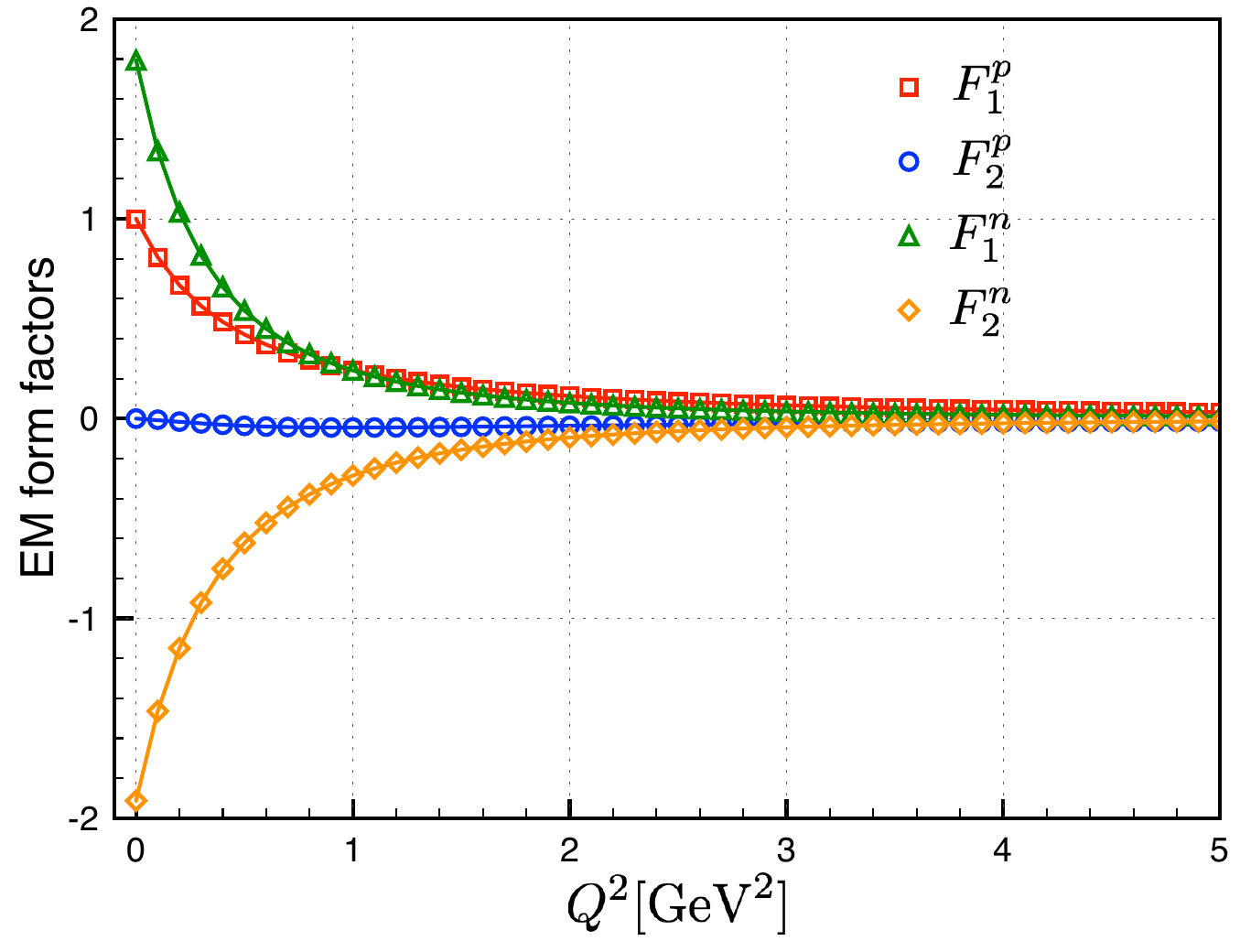}}{-0.4cm}{0.0cm}
\topinset{(b)}{\includegraphics[width=7.5cm]{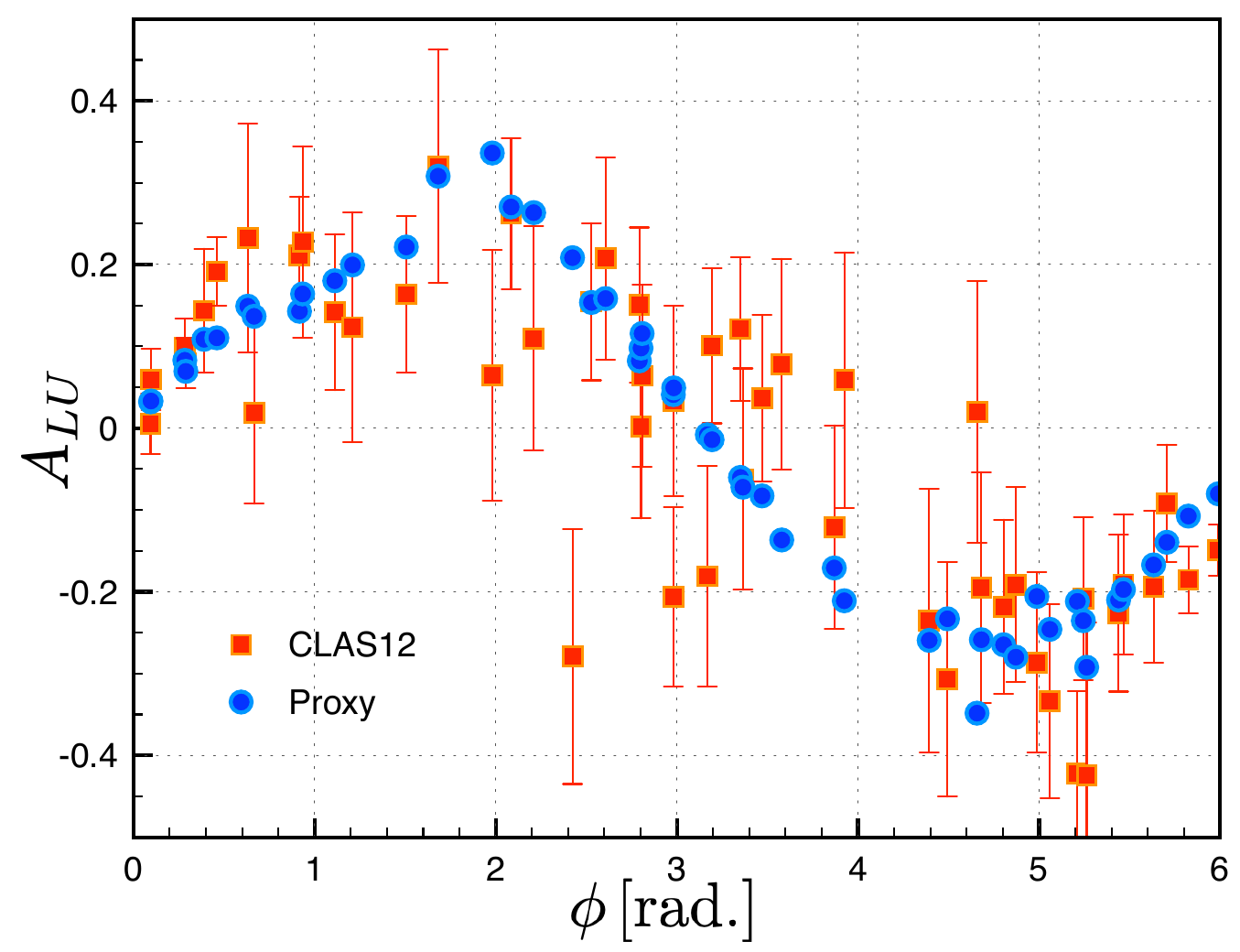}}{-0.4cm}{0.0cm}
\caption{(Color online) Consistency checks for the present reduced-profile MEM fit.  (a) Electromagnetic form factors $F_{1,2}^{p,n}$ as functions of $Q^2$.  The symbols denote the input form factors, while the solid curves are reconstructed from the GPD moments.  (b) Qualitative comparison between the CLAS12 beam-spin-asymmetry data of Ref.~\cite{CLAS:2022syx} and the GPD-based proxy $A_{LU}^{\rm proxy}=K\,{\rm Im}\,\mathcal C_{\rm unp}^I\sin\phi$.  Panel (b) illustrates the leading $\sin\phi$-type behavior generated by the reconstructed profiles.  This comparison is not used as a constraint on the GPD profiles and is meant as a qualitative check of the azimuthal pattern.}
\label{FIG2}
\end{figure}

Fig.~\ref{FIG2} summarizes two consistency checks of the present construction. Fig.~\ref{FIG2}(a) compares the input electromagnetic form factors with those reconstructed from the first moments of the MEM-generated GPDs as already demonstrated in Table~\ref {tab:input_vs_gpd_form_factors}. The close agreement between the symbols and curves confirms that the reduced-profile solution reproduces the imposed elastic form-factor constraints within the accuracy of the present baseline calculation. Fig.~\ref{FIG2}(b) compares the CLAS12 beam-spin-asymmetry data with the simple GPD-based proxy and shows that the reconstructed profiles are compatible with the expected leading $\sin\phi$ pattern.

For this comparison, the GPD values are
interpolated at
\begin{equation}
x=\xi=\frac{x_B}{2-x_B},
\qquad
t=t_{\rm exp},
\end{equation}
for each experimental point. The resulting theoretical curve contains the point-by-point kinematic dependence of ${\rm Im}\mathcal C_{\rm unp}^{I}(x_B,t)$ in addition to the leading $\sin\phi$ dependence.  A full DVCS cross-section calculation would require the exact Bethe-Heitler lepton propagators, the full $\phi$-dependent denominator, the real parts of the CFFs, polarized GPDs such as $\widetilde{H}$, and possible higher-twist contributions.  The fitted factor $K$ therefore absorbs the dominant kinematic prefactors and part of the missing $\phi$-dependent structure.  The values in Table~\ref{tab:dvcs_proxy_fit} show a stable overall normalization across the tested beam-energy subsets, while the corresponding $\chi^2/{\rm dof}$ values quantify the limited scope of the proxy.

%-----------------------------------------------------------
\subsection{Systematic uncertainties of the present reconstruction}
\label{subsec:systematics}
%-----------------------------------------------------------
The present construction is a constrained zero-skewness baseline solution.  The elastic form factors provide $x$-integrated constraints, and the $x$-dependent transverse information is obtained after introducing forward inputs, a reduced positive profile ansatz, and entropy regularization.  We therefore summarize here the main systematic limitations of the calculation and describe how they can be quantified.

First, the down-quark Dirac profile $f_H^d(x)$ is the profile most susceptible to prior dependence in less-restricted reconstructions, an expectation that follows directly from the flavor structure of the electromagnetic current: the $d$-quark contribution to the proton form factors carries the smaller charge weight $-1/3$, and the large-$x$ behavior of
$d_v(x)$ is more strongly suppressed than that of $u_v(x)$.  Hence, the elastic moment constraints determine the integrated $d$-quark contribution more strongly than the local shape of $f_H^d(x)$, especially at large $x$.  The corresponding uncertainty can be propagated to any derived observable ${\cal O}$ by evaluating
\begin{equation}
\Delta_{\alpha}{\cal O}
=
\frac{1}{2}
\left[
\max_{\alpha}{\cal O}(\alpha)
-
\min_{\alpha}{\cal O}(\alpha)
\right],
\end{equation}
over the tested range of $\alpha$.  For example, the $d$-quark Dirac
transverse size is
\begin{equation}
\langle b_\perp^2\rangle_{H_d}(\alpha)=\frac{\int_0^1 dx\,d_v(x)\,4f_H^d(x;\alpha)}
{\int_0^1 dx\,d_v(x)}.
\end{equation}
The spread of this quantity gives a direct estimate of how much the down-quark transverse size depends on the entropy weight.

Second, the beam-spin-asymmetry proxy carries the expected limitations of a zero-skewness input.  The approximation $H_v^q(\xi,\xi,t)\simeq H_v^q(\xi,0,t)$ and $E_v^q(\xi,\xi,t)\simeq E_v^q(\xi,0,t)$ neglects the genuine skewness dependence of the GPDs, while the effective parameter $K$ in Eq.~(\ref{eq:ALU}) absorbs the Bethe-Heitler-dominated denominator, exact kinematic prefactors, and missing $\phi$-dependent terms.  A practical estimate of this uncertainty can be obtained by fitting $K$ separately in kinematic bins, comparing point-by-point kinematics with a fixed-bin-center kinematic approximation, and using the resulting spread as a systematic uncertainty of the diagnostic comparison.

Third, the forward Pauli input $e_v^q(x)$ introduces a model uncertainty in the Pauli sector.  Only the first moments of $e_v^q$ are fixed by $\kappa_q$, while the detailed $x$-shape is parametrized.  Varying $(a_E^q,b_E^q)$ at fixed $\kappa_q$ changes the $x$-weighting of the Pauli form-factor constraints and can modify $f_E^q(x)$, $E_v^q(x,t)$, and derived quantities such as $\langle b_\perp^2\rangle_{E_q}$.  Such sensitivity should be folded into a Pauli-input systematic uncertainty in future fits.

Finally, we stress that the sign of $E_v^d$ is not an additional systematic effect: it is fixed by $\kappa_d<0$ and was already built into the entropy treatment of Sec.~\ref{sec:mem_formulation}, so the corresponding impact-parameter-space distribution remains a signed, not positive-definite, quantity throughout.

%--------------------------------------------------
\section{Summary}
\label{sec:summary}
%--------------------------------------------------
Starting from the four nucleon electromagnetic form factors $F_1^p$, $F_1^n$, $F_2^p$, and $F_2^n$, converted into flavor constraints on $F_1^u$, $F_1^d$, $F_2^u$, and $F_2^d$, we have shown that a low-dimensional, positive transverse profile can be pinned down once these elastic moments are combined with the forward-PDF and anomalous-moment normalizations through the ansatz $H_v^q(x,t)=q_v(x)\exp[t f_H^q(x)]$ and $E_v^q(x,t)=e_v^q(x)\exp[t f_E^q(x)]$.  Maximizing an entropy functional subject to these simultaneous constraints determines the unknown profiles $f_H^q(x)$ and $f_E^q(x)$.

A key point of the present work is the reduction of profile degrees of freedom.  A direct grid-based profile contains many local modes that are weakly constrained by elastic form factors, because form factors provide only $x$-integrated moment information.  We therefore use the positive three-parameter profile $f(x)=0.05+(1-x)^2\exp(c_0+c_1x+c_2x^2)$, reducing the number of profile parameters from $4N_x$ to twelve in the present four-profile analysis while preserving the expected large-$x$ suppression and stabilizing the reconstruction against variations of the entropy weight.

The numerical checks show that the resulting profiles reproduce the elastic form-factor input through the GPD moments, satisfy the forward-limit PDF and anomalous-moment normalizations, and remain stable under variations of $\alpha$ over $10^{-4}\le\alpha\le10^{-2}$.  The resulting profiles also allow a direct impact-parameter-space interpretation. The transverse distributions show the expected shrinkage as $x$ increases, and the signed Pauli-type distributions reflect the flavor-anomalous magnetic moments.  The beam-spin-asymmetry proxy gives a complementary qualitative check of the leading $\sin\phi$-type pattern, while its $\chi^2/{\rm dof}$ values indicate the expected room for improvement in a complete DVCS treatment.

The present construction uses the lowest elastic moments together with forward inputs and a reduced positive-profile manifold to determine the $x$-dependent transverse profiles.  These choices enter as interchangeable modules rather than as intrinsic assumptions of the method: the coupled MEM pipeline formulated here accommodates alternative elastic parametrizations, modern PDF sets, and future data-driven constraints without structural change.  Future work should replace the controlled elastic input with global form-factor fits including covariance information, propagate PDF and Pauli-input uncertainties, include lattice-QCD generalized form factors, test default-model dependence, and extend the method to polarized and nonzero-skewness GPDs using DVCS and other hard exclusive observables.  These extensions should promote the present reduced-profile MEM construction toward a quantitative phenomenological extraction.

%--------------------------------------------------
\section*{Acknowledgment}
%--------------------------------------------------
This work was supported by the National Research Foundation of Korea (NRF) grant funded by the Korean government (MSIT) under Grant No. NRF-RS-2025-16065906.
%-------------------------------------------------

%-------------------------------------------------

\begin{thebibliography}{99}
%-------------------------------------------------
\bibitem{Muller:1994ses}
D.~M{\"u}ller, D.~Robaschik, B.~Geyer, F.~M.~Dittes and J.~Ho{\v{r}}ej{\v{s}}i,
%``Wave functions, evolution equations and evolution kernels from light ray operators of QCD,''
Fortsch. Phys. \textbf{42}, 101-141 (1994).
%doi:10.1002/prop.2190420202.
%[arXiv:hep-ph/9812448 [hep-ph]].
%-------------------------------------------------
\bibitem{Ji:1996ek}
X.~D.~Ji,
%``Gauge-Invariant Decomposition of Nucleon Spin,''
Phys. Rev. Lett. \textbf{78}, 610-613 (1997).
%doi:10.1103/PhysRevLett.78.610
%[arXiv:hep-ph/9603249 [hep-ph]].
%-------------------------------------------------
\bibitem{Radyushkin:1996nd}
A.~V.~Radyushkin,
%``Scaling limit of deeply virtual Compton scattering,''
Phys. Lett. B \textbf{380}, 417-425 (1996).
%doi:10.1016/0370-2693(96)00528-X
%[arXiv:hep-ph/9604317 [hep-ph]].
%-------------------------------------------------
\bibitem{Radyushkin:1997ki}
A.~V.~Radyushkin,
%``Nonforward parton distributions,''
Phys. Rev. D \textbf{56}, 5524-5557 (1997).
%doi:10.1103/PhysRevD.56.5524
%[arXiv:hep-ph/9704207 [hep-ph]].
%-------------------------------------------------
\bibitem{Goeke:2001tz}
K.~Goeke, M.~V.~Polyakov and M.~Vanderhaeghen,
%``Hard exclusive reactions and the structure of hadrons,''
Prog. Part. Nucl. Phys. \textbf{47}, 401-515 (2001).
%doi:10.1016/S0146-6410(01)00158-2
%[arXiv:hep-ph/0106012 [hep-ph]].
%-------------------------------------------------
\bibitem{Diehl:2003ny}
M.~Diehl,
%``Generalized parton distributions,''
Phys. Rept. \textbf{388}, 41-277 (2003).
%doi:10.1016/j.physrep.2003.08.002
%[arXiv:hep-ph/0307382 [hep-ph]].
%-------------------------------------------------
\bibitem{Guidal:2013rya}
M.~Guidal, H.~Moutarde and M.~Vanderhaeghen,
%``Generalized Parton Distributions in the valence region from Deeply Virtual Compton Scattering,''
Rept. Prog. Phys. \textbf{76}, 066202 (2013).
%doi:10.1088/0034-4885/76/6/066202
%[arXiv:1303.6600 [hep-ph]].
%-------------------------------------------------
\bibitem{Kumericki:2016ehc}
K.~Kumericki, S.~Liuti and H.~Moutarde,
%``GPD phenomenology and DVCS fitting: Entering the high-precision era,''
Eur. Phys. J. A \textbf{52}, no.6, 157 (2016).
%doi:10.1140/epja/i2016-16157-3
%[arXiv:1602.02763 [hep-ph]].
%-------------------------------------------------
\bibitem{Kroll:2012sm}
P.~Kroll, H.~Moutarde and F.~Sabatie,
%``From hard exclusive meson electroproduction to deeply virtual Compton scattering,''
Eur. Phys. J. C \textbf{73}, no.1, 2278 (2013).
%doi:10.1140/epjc/s10052-013-2278-0
%[arXiv:1210.6975 [hep-ph]].
%-------------------------------------------------
\bibitem{Burkardt:2000za}
M.~Burkardt,
%``Impact parameter dependent parton distributions and off forward parton distributions for zeta ---{\ensuremath{>}} 0,''
Phys. Rev. D \textbf{62}, 071503 (2000)
[erratum: Phys. Rev. D \textbf{66}, 119903 (2002)].
%doi:10.1103/PhysRevD.62.071503
%[arXiv:hep-ph/0005108 [hep-ph]].
%-------------------------------------------------
\bibitem{Skilling:1984}
J. Skilling,
%Data analysis: The maximum entropy method. 
Nature 309, 748-749 (1984). 
%https://doi.org/10.1038/309748a0
%-------------------------------------------------
\bibitem{Jarrell:1996rrw}
M.~Jarrell and J.~E.~Gubernatis,
%``Bayesian inference and the analytic continuation of imaginary-time quantum Monte Carlo data,''
Phys. Rept. \textbf{269}, 133-195 (1996).
%doi:10.1016/0370-1573(95)00074-7
%-------------------------------------------------
\bibitem{Kelly:2004hm}
J.~J.~Kelly,
%``Simple parametrization of nucleon form factors,''
Phys. Rev. C \textbf{70}, 068202 (2004).
%doi:10.1103/PhysRevC.70.068202
%-------------------------------------------------
\bibitem{Hill:2010yb}
R.~J.~Hill and G.~Paz,
%``Model independent extraction of the proton charge radius from electron scattering,''
Phys. Rev. D \textbf{82}, 113005 (2010).
%doi:10.1103/PhysRevD.82.113005
%[arXiv:1008.4619 [hep-ph]].
%-------------------------------------------------
\bibitem{Ye:2017gyb}
Z.~Ye, J.~Arrington, R.~J.~Hill and G.~Lee,
%``Proton and Neutron Electromagnetic Form Factors and Uncertainties,''
Phys. Lett. B \textbf{777}, 8-15 (2018).
%doi:10.1016/j.physletb.2017.11.023
%[arXiv:1707.09063 [nucl-ex]].
%-------------------------------------------------
\bibitem{Martin:2009iq}
A.~D.~Martin, W.~J.~Stirling, R.~S.~Thorne and G.~Watt,
%``Parton distributions for the LHC,''
Eur. Phys. J. C \textbf{63}, 189-285 (2009).
%doi:10.1140/epjc/s10052-009-1072-5
%[arXiv:0901.0002 [hep-ph]].
%-------------------------------------------------
\bibitem{Owens:2012bv}
J.~F.~Owens, A.~Accardi and W.~Melnitchouk,
%``Global parton distributions with nuclear and finite-$Q^2$ corrections,''
Phys. Rev. D \textbf{87}, no.9, 094012 (2013).
%doi:10.1103/PhysRevD.87.094012
%[arXiv:1212.1702 [hep-ph]].
%-------------------------------------------------
\bibitem{Dulat:2015mca}
S.~Dulat \textit{et al.}, 
%T.~J.~Hou, J.~Gao, M.~Guzzi, J.~Huston, P.~Nadolsky, J.~Pumplin, C.~Schmidt, D.~Stump and C.~P.~Yuan,
%``New parton distribution functions from a global analysis of quantum chromodynamics,''
Phys. Rev. D \textbf{93}, no.3, 033006 (2016).
%doi:10.1103/PhysRevD.93.033006
%[arXiv:1506.07443 [hep-ph]].
%-------------------------------------------------
\bibitem{Vanderhaeghen:1999xj}
M.~Vanderhaeghen, P.~A.~M.~Guichon and M.~Guidal,
%``Deeply virtual electroproduction of photons and mesons on the nucleon: Leading order amplitudes and power corrections,''
Phys. Rev. D \textbf{60}, 094017 (1999).
%doi:10.1103/PhysRevD.60.094017
%[arXiv:hep-ph/9905372 [hep-ph]].
%-------------------------------------------------
\bibitem{CLAS:2022syx}
G.~Christiaens \textit{et al.} [CLAS],
%``First CLAS12 Measurement of Deeply Virtual Compton Scattering Beam-Spin Asymmetries in the Extended Valence Region,''
Phys. Rev. Lett. \textbf{130}, no.21, 211902 (2023).
%doi:10.1103/PhysRevLett.130.211902
%[arXiv:2211.11274 [hep-ex]].
%-------------------------------------------------
\end{thebibliography}
\end{document}